# Democracy, essential element of the electronic government

Bostan I.

**Abstract**— This paper emphasizes a determinant aim of identifying different approaches, as comparing to the education and democracy ways specific to e-government system. Introducing the information technology should offer the possibility by which reform processes of the government should become more efficient, transparent and much more public for the citizens; in this way, their ability of participating directly to government activities should prove the carrying out of a democratic and free frame.

One of the essential issues of such phenomenon is that of proving that adopting the information and communication technology programs to government process or electronic government depends upon a series of external factors, such as the level of state's development, the cultural level, the frame of developing the structures of central and local public authority, criteria that differentiate the applicability of such system, to various countries. This difference is especially seen as comparing to the East states of European Union. Information systems can be applied in order to allow the citizens to monitor and coordinate the providing of local services; such exchanges have created trust and the feeling of influence, encouraging the participation to political life. Carrying into effect the new informational technologies, aiming to issuing, informing and to participation of citizens to political life, will model the concept of democracy within a new frame.

—————— ◆ ——————

## 1 INTRODUCTION

The minimal rate of participation to political life and especially as comparing to different elections coming from European electros, the stage of general estrangement as result and the level of not being interested have brought the creation of a form with "democratic deficit", which influenced most of European Union states in the last years. The governments have seen in this low participation a threatening to civil society address, fact that generated the initiation of some significant investments, so as to improve the new informational and communication technologies, and in order to introduce the programs, projects so as to stimulate the activity of citizens in this field. Most of these e-government proposals have involved an improved access of citizens to services. These projects have offered exact benefits, as regards the administrative efficiency; in this way, the citizens have improved the access to resources or to provide resources. Citizens, as services consumers, have won regarding the activity of administrative bureaucracy, but few of such projects have increased to the level of citizens' participation to processes of decision taking. The e-government process is an information system of managing the governmental politics, which is not necessarily dealing with the democratic deficit, where the system is just an informing adjutant.

A particular attention was given to the issuance of state's politics, as concerns the labor conscription, where this thing was described as "e-government" (according to Pierre 2000, for a general discussion of government)[1]. In this way, the policy of EU has tried to increase the democratic participation, by using the technology so as to facilitate the information technology as comparing to civic society, by including the dialogue – online forums, discussion by virtual cams or electronic elections. There are few examples of such projects on significant impact, and few proofs of some new developing formal structures, able to encourage the citizens so as to effectively participate towards forming the politics. The projects will receive direct support from European Union, and a recent politics document, "The role of e-government to Europe's future" (Commission of European Communities, 2003)[2], might propose that e-government is represented by strategies that should "promote on-line the democratic participation". It seems that EU politics was efficient so as to ameliorate the efficiency, but does not improve the participation of citizens to the program of e-government, meaning that this might dispose of a decisional activity towards the politics of its state.

## 2. CRITICAL FACTORS OF INFLUENCE

The full and effective participation to information network in progress of development at world level has a fundamental importance for a country that aims on

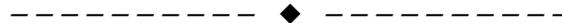

- *Bostan I. is from the University of Suceava, 13 University Str., Suceava, Romania.*

---

[1] Pierre, J. (Ed.). (2000). Debating Governance: Authority, Steering and Democracy. Oxford: Oxford University Press.
[2] Commission of the European Communities. (2003). The Role of e-Government for Europe's Future (SEC(2003) 1038). Brussels: Commission of the European Communities.

avoiding the marginalization of globalization process, and has also become essential for full participation of its citizens in all spheres of life. Information and communication technology (ICT) can contribute on integrating the countries in progress of development to world economy and can create the conditions of information and exchange of knowledge and using. ICT offers a huge potential in order to improve the living standards and opportunities for individuals, communities of various regions. While many states of the world have remained directly not reached by the informational revolution, the effect of transformation cannot be neglected, as result over the global society.

The governments are facing a new reality of the imperatives changed as result of spreading ICT within the entire world. This thing needs a fundamental changing, meaning the way the state acts on internal level and interacts with its citizens, especially in its function of promoting good governance, as condition to lasting development. Occurrence of informational society represents the creation of unprecedented conditions, so as to carry out this function. By means of ICT applications and e-government, the communication between the administration, the citizens and the enterprises can be improved so as to perfecting the governance and management of public sector, the access to economic and social opportunities, as well as the reducing of digital discrepancy into a society.

Conventional using of "e" prefix suggests that an activity is "electronic" or of digital nature. By accepting this fact, the term of e-governance might refer to only utilization of electronic information and of communication technologies of communication in the enterprise of governmental activities, in education, health, agriculture, government, customs, etc.

The e-governance should be considered as an alternative or complementary approach over the governmental management or delivering the services, as means of redefining the manner on how the citizens interact of private sector with the public sector of administration. In this way:

- The governments should use ICT in order to minimize the transaction costs and to simplify the procedures of bureaucracy, fact that makes the accomplishment of operation more efficient, issuing resources able to allow services within better organized and economic way.
- Governments should achieve better results, so as to accomplish the objectives of development, using ICT that bring towards the increasing of relevance specific to process of formulating the politics, by participating to needs of citizens and by increasing the quality of their services.
- "e" signifies the responsibility: ICT can support the increasing of interaction between citizens and their governments, both for the citizens, by participating on processes of taking decisions and by their awareness to community's development.
- "e" signifies the economic and social development: beyond the economic benefits, which are accumulated on governance, due to the efficiency and advantages, the using of ICT on governments and on the interaction with business community and citizens might create new levers on attracting the investments, as well as creating new employment opportunities.

As result of those above mentioned, the interaction between government and citizens is established depending upon the priorities by which this allows the participation to political life. There are politics instruments by which the governments can increase the civic participation. In this way, the new technologies can have a significant part. There are many reasons as regards the lack of civic participation within politics or administration. One of the causes can be represented by the lack of information, as regards the political decisions that are about to be performed and a lack of information, concerning the way of participation, as citizen, on taking such decisions. Another cause might be the mistrust towards the impartiality and correctness of these decisions, so that citizens do not believe their interventions as efficient. Although, another cause can consists in the lack of wish, concerning the wish of entering within a political topic; if people are contended by the existing system, they do not see any reason for a participation, in these situations, the lack conviction specific to entering into politics might have an impact, as well as a lack of knowledge about the way of transmitting their opinions, that brings towards on estrangement or disinformation toward the political process. In this way, the decreased levels of trust and social solidarity represent a reason of reducing the level of political participation (Putnam, 2000; Putnam, Leonardi, & Nanetti, 1993).[3]

Notwithstanding, the means by which the social trust, the mutual feature, the solidarity can be increased are not too clear, and no connection between trust and solidarity exists, by which the political participation gained development. New technologies of information and communications have been promoted as mechanisms, by which the social formation of community can be improved, with benefits on long term to civil society. One of the first experiments in using the technologies so as to increase the public participation was in Santa Monica, in 1989 (Docter & Dutton, 1998)[4], was one of experiments on building new technologies (Tsagarousianou, Tambini, si Bryan , 1998). To more of such studies, only a small percentage of local people used the technology, and it became difficult to create extrapolations about technology, in community and in participation to political life. The plan of activity as concerns the e-government for development is the result of an ample consultative and process, which involves representatives of the states in progress of development, private sector, non-governmental organizations, academic institutions and international organizations. This plan of action aims on providing an orientation and a platform of governmental or non-governmental institutions and international organizations, which participate on development and supporting the ability necessary as public service, on becoming more efficient.

---

[3] Putnam, RD (2000). Bowling alone: America's declining social capital. New York: Simon & Schuster. Putnam, RD, Leonardi, R., & Nanetti, R. (1993). Making democracy work: civic traditions in modern Italy. Princeton, NJ: Princeton University Press.

[4] Docter, S., & Dutton, WH (1998). The First Amendment online: Santa Monica's public electronic network. In R. Tsagarousianou & D. Tambini & C. Bryan (Eds.), Cyberdemocracy: technology, cities and civic networks (pp. 125-151). Londra: Routledge.

Extending the institutional ability within sector of politics development, the reform of public sector, the legal and regulation frame, strategic planning and management of changing, as well as coordinating the intergovernmental relationships will be necessary so as to harmonize the process concerning the transformation and exchange of information between different entities that compose the system of government. The initial steps that were identified so as to direct states, where the e-government project is applied, are represented by:

- Drawing up an environment analysis (key factors towards the e-government)
- Issuing the long term visions, including the contribution on developing and solutions on waiting, foreseen by applying the e-government project
- Drawing up the aimed strategic objectives
- Identifying the priorities and the expected impact

## 3. KEY FACTORS CONCERNING THE E-GOVERNMENT IMPLEMENTATION

Implementing the e-government project needs a proper environment so as to maximize its potential. Before defining the e-government, knowing the strategy of development or an action plan is necessary, as regards a throughout analysis of the existing environment, where the e-government project will be applied. The government can ask few questions so as to evaluate the way of implementing the informational program to e-government, and so as to evaluate the level of requesting and of citizens' involvement specific to this project.

| Areas | Key factors |
|---|---|
| Level of political development of the state | - Awareness over the political value and of level of e government
- Commitment towards the e-government and good government
- Abilities on management as regards the class of government
- National identity and perception of democratic government
- Legislative frame
- Activity of citizens and participation of civil society in actions performed by the government |
| Frame of settlement (for creating the economic conditions to accessible ICT, infrastructures, services and equipments are necessary) | - Security standards
- Legislation privacy
- Judicial validity of on-line transactions
- Level of freedom specific to telecommunications market, of internet services on the market
- Positive fiscal environment, for the acquisition of IT equipments |
| Conditions of organization | - Decentralized administrative structures
- Public administration reforms
- Central coordination and supporting units in the field
- Coordination of government politics on ICT system
- Governmental inter-relationships |
| Cultural conditions and human resources | - Culture, traditions and languages
- Levels of education
- IT alphabetization and the number of online users
- IT in institutions of education and of programs
- Cultures of information and exchange of knowledge
- Organizational culture predominance
- Attitude and adaptability on changing, especially on public administration
- Managerial abilities in public sector
- Service of orientation on public administration towards citizens |
| Financial conditions (initial costs related to applying the e-government) | - Process of resources allocation
- Structure of national income
- Access to alternative mechanism of financing
- Partnerships with private |

|  |  |
|---|---|
|  | sector<br>• Access to capital markets<br>• Mechanisms for risk investments |
| **Technological infrastructure** (the lack of technologies is a major problem for countries aiming to apply and maintain the e-government) | • (Tele) infrastructure of communications<br>• Rates of involving the telecommunications services<br>• Urban towards rural: demographical/geographical prejudice<br>• Software and hardware (inherited systems)<br>• IT standards |
| **Data and information systems** (management systems, official deeds and work processes, which have to provide necessary data on supporting the adopting of e-government) | • Data available and accessible of information<br>• Procedures of collecting the data and of storing the information by standardization<br>• Quality of data and information, accomplishing the secure process of protecting data<br>• Ability of analyzing the data and of dissimulating the using of information<br>• Ability of implementing the flows of information directly within decisional processes<br>• Legislative, political information and the impact towards the citizen |

Starting from the evaluation of these key factors, one might notice that the access to political life by a citizen, or his involvement in taking decisions towards the government politics, can be accomplished pretty difficult, existing some limits in this way, determined by external factors. Strategic objectives as regards the implementation of such system consist in making sensitive the public opinion, so that the civic society within the frame of a democratic state should participate actively on understanding the government situations and not only in the situation of a spectator. An important part that Government has, in this way, consists in participation and consulting the decisional processes within private sector, civic society, by providing online services and Internet.

## CONCLUSIONS

E-government aims on reaching multi-dimensional limits, able to be found especially on economic and social levels, as well within process of political government. The process of priorities has to be focused over effects in central perspective towards a perspective directed towards development.

These programs have sometimes included projects that use new technologies, with the aim of increasing the participation of public on formulating the politics of encouraging the development of volunteering groups. The trying of increasing the participation to volunteering groups had a limited success, and research suggested that new technologies do not increase the number of participants. Notwithstanding, the new technologies do not increase the level of informational communication, but distributing a controlled information, by which a governmental transparency is created. The phenomenon brings into public opinion a positive nuance, since the access to ICT systems by citizens combined with an increased responsibility will make possible their intervention on monitoring the activities of governance and implicitly, the impact of these will bring to continuity of participation, so as to create the democratic frame.

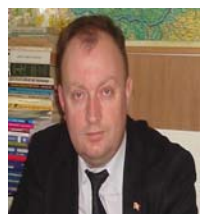
**Ionel Bostan, Prof. PhD** - Stefan cel Mare University, authors 5 speciality books, over 10 papers published in journals rated by ISI Thompson and 30 scientifically papers published in the country and abroad at the International Symposiums or Conferences. He is a member in 10 international professional organizations and scientifically referee in the editing committee of 2 journals rated by REPEC, Socionet, Research Gate etc, and within the scientifically committee of the WASET.